\documentclass[12pt]{spieman}  

\usepackage{amsmath,amsfonts,amssymb}
\usepackage{rotating}
\usepackage{hyperref}
\usepackage{setspace}
\usepackage{tocloft}
\usepackage{enumitem}
\usepackage{ltablex}
\usepackage{graphicx}
\usepackage{float}
\usepackage{threeparttable}
\usepackage{hhline}
\usepackage{lineno}
\usepackage{booktabs}
\usepackage[table]{xcolor}
\usepackage{booktabs,tabulary}
\usepackage{siunitx}

\usepackage[normalem]{ulem}
\title{Studying the Impact of Optical Aberrations on Diffraction-Limited Radial Velocity Instruments} 

\author[a]{Eric B. Bechter}
\author[a]{Andrew J. Bechter}
\author[a]{Justin R. Crepp}
\author[a]{Jonathan Crass}

\affil[a]{University of Notre Dame, Department of Physics, 225 Nieuwland Science Hall, Notre Dame, IN 46556, USA}

\cftpagenumbersoff{figure}
\cftpagenumbersoff{table} 
\begin{document} 
\maketitle

\begin{abstract}
Spectrographs nominally contain a degree of quasi-static optical aberrations resulting from the quality of manufactured component surfaces, imperfect alignment, design residuals, thermal effects, and other other associated phenomena involved in the design and construction process. Aberrations that change over time can mimic the line centroid motion of a Doppler shift, introducing radial velocity (RV) uncertainty that increases time-series variability. Even when instrument drifts are tracked using a precise wavelength calibration source, barycentric motion of the Earth leads to a wavelength shift of stellar light causing a translation of the spectrum across the focal plane array by many pixels. The wavelength shift allows absorption lines to experience different optical propagation paths and aberrations over observing epochs. We use physical optics propagation simulations to study the impact of aberrations on precise Doppler measurements made by diffraction-limited, high-resolution spectrographs. Using the optical model of the \textit{iLocater} spectrograph, we quantify the uncertainties that cross-correlation techniques introduce in the presence of aberrations and barycentric RV shifts. We find that aberrations which shift the point-spread-function (PSF) photo-center in the dispersion direction, in particular primary horizontal coma and trefoil, are the most concerning. To maintain aberration-induced RV errors less than 10 cm/s, phase errors for these particular aberrations must be held well below 0.05 waves at the instrument operating wavelength. Our simulations further show that wavelength calibration only partially compensates for instrumental drifts, owing to a behavioural difference between how cross-correlation techniques handle aberrations between starlight versus calibration light. Identifying subtle physical effects that influence RV errors will help ensure that diffraction-limited planet-finding spectrographs are able to reach their full scientific potential.
\end{abstract}

{\noindent \footnotesize\textbf{Address all correspondence to:} E. Bechter  
\\ Email:\linkable{ebechter@nd.edu}\\ 
Accepted to JATIS}

\begin{spacing}{1.5}   

\section{Introduction}\label{sec:intro}

The point-spread-function (PSF) of a spectrograph illuminated using a diffraction-limited source describes the intensity broadening introduced by diffraction from optical components within the system \cite{Figueira_18}. An ideal spectrograph would offer end-to-end diffraction-limited performance. However, optical aberrations act to spatially redistribute spectral energy, reducing both spectral resolution and local signal-to-noise ratio. All optical components, whether refractive or reflective, impart some level of wavefront error which can modify the instrument PSF as a function of the 2D field and consequently the measured spectrum across the detector. Even in the case of a theoretically idealized spectrograph, mechanical flexure, thermal variations, and systematic misalignment further introduce aberrations. Without proper consideration and modeling, these real-world effects can contribute non-negligibly to the error budgets of extremely precise radial velocity (EPRV) instruments, particularly those that operate at the diffraction-limit fed by single-mode fibers (SMF)\cite{Gibson_16}.

Wavelength calibration units are used to track the location of specific wavelengths of light that reach the detector focal plane. The warping and distortion of image locations across the focal plane is provided as a natural by-product of establishing a wavelength solution, e.g. through the use of etalon or laser frequency comb peaks that result from the periodic constructive interference of light as a function of wavelength. However, higher-order effects that modify the shape of the incident PSF onto the detector are not necessarily captured in the wavelength calibration process; calibration data may be recorded non-contemporaneously, or in different field locations using different fibers from a science target being observed. Additionally, ideally the PSF profile should be wavelength independent, however, even if aberrations are stable in time and geometrically similar across the array, the effective \emph{strength} of the aberrations is inherently chromatic. Since image quality degrades based on the amplitude of phase errors compared to the wavelength of light, the shortest wavelengths will experience reduced Strehl ratios relative to longer wavelengths. For instruments with a broad bandpass, only a fraction of the spectrum may be truly diffraction-limited.

Further considerations are time-varying Doppler shifts effects. Due to the Earth's barycentric motion, spectra recorded at different epochs will be Doppler shifted and spectral features will be recorded at different locations on the detector. At the resolution of planet-finding spectrographs ($R>100,000$), a barycentric motion of $\pm 30$ km s$^{-1}$ corresponds to a translation across dozens of pixels. Over an observing season, measurements of the same spectral features will be subject to different optical paths and therefore varying field aberrations. As next-generation spectrographs are required to offer broad wavelength coverage and high spectral resolution to address stellar variability at sub-meter-per-second precision level (e.g. EXPRES\cite{Jurgenson_16}, NEID\cite{Schwab_16}, MAROON-X\cite{Seifahrt_16}, ESPRESSO\cite{Pepe_10}, G-CLEF\cite{Szentgyorgyi_14}, and iLocater\cite{Crepp_14}; see \citenum{Plavchan_15} and \citenum{Fischer_16} for a comprehensive list), the translation of spectral features can be significant. Large format arrays used by these instruments require significant fields of view which can exacerbate the challenge to maintain image quality over off-axis field angles. As mitigating stellar variability requires tracking higher-order features of the stellar spectrum, such as absorption line asymmetries, making wavelength-dependent image quality and calibration is a crucial instrument feature.

\begin{figure*}[t]
    \centering
    \includegraphics[trim=4.0cm 1.0cm 4.0cm 2.0cm,clip,width=0.8\textwidth]{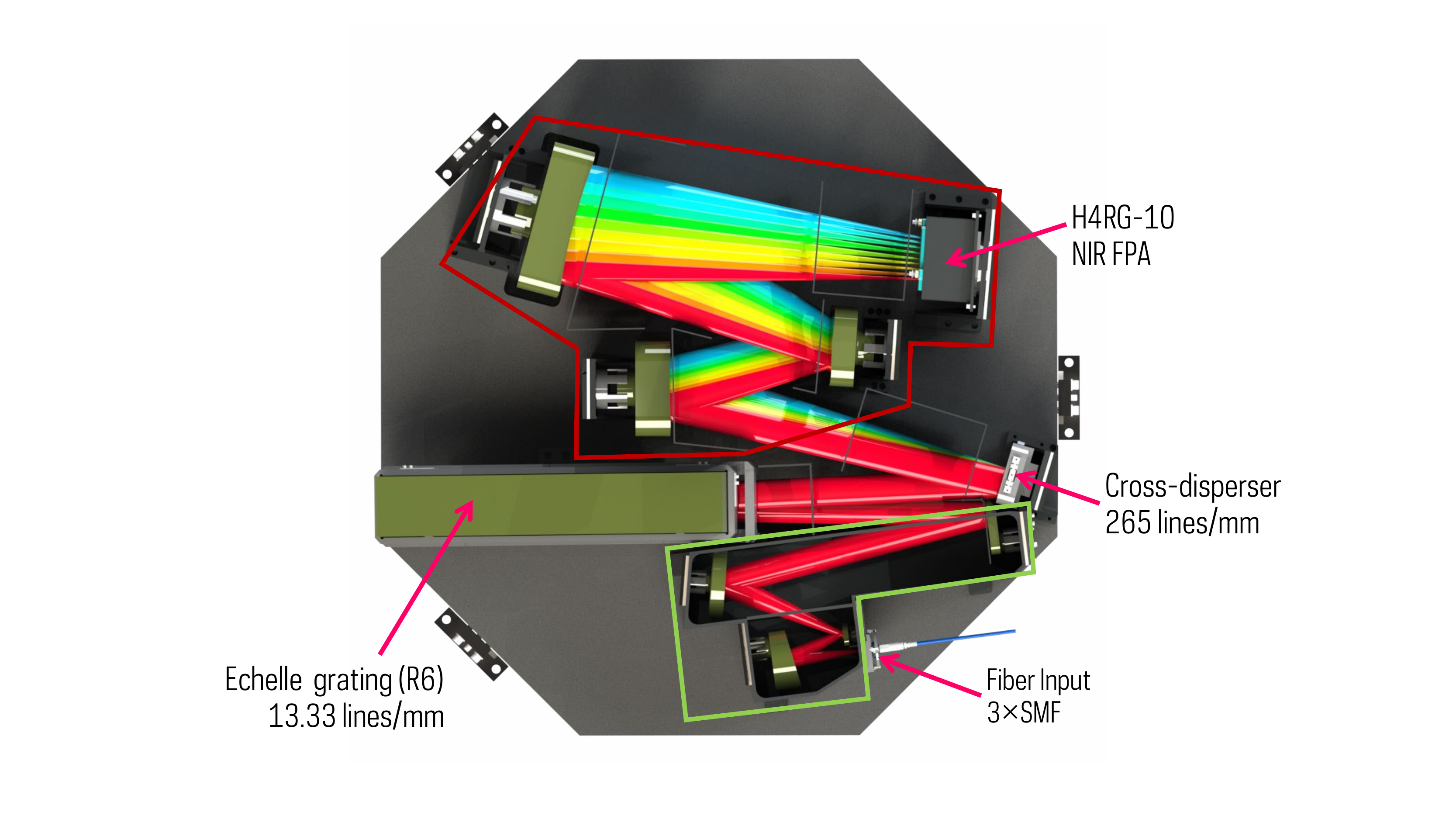}
    \caption[The iLocater spectrograph.]{CAD rendering of the iLocater spectrograph. The spectrograph is illuminated using three single-mode fibers (SM980) which are collimated onto an R6 echelle grating. The beam is cross dispersed using a reflection grating and imaged onto an H4RG-10 near-infrared focal plane array (FPA). The collimator (green) and camera (red) have effective focal lengths of 113.4mm and 440mm respectively giving an end-to-end system magnification of 3.9. The system has been designed to maintain end-to-end diffraction-limited performance.}
    \label{fig:ilocater_spectrograph}
\end{figure*}

As spectrograph error budgets and opto-mechanical tolerances continue to tighten to meet scientific needs, RV instrument designs and reduction pipelines must consider the impact of aberrations on data products. In this paper, we explore the impact of optical aberrations on diffraction-limited planet-finding spectrographs by studying focal plane interactions between stellar spectra, Doppler shifts, barycentric motion, and wavelength calibration. The optical prescription for the iLocater spectrograph (Figure~\ref{fig:ilocater_spectrograph}), which is being developed for the Large Binocular Telescope (LBT) in Arizona, is used as a baseline design. Simulations and methodology details, including how aberrations are generated and quantified as RV errors, are described in \S\ref{sec:method}. Results for: (i) spectral resolution degradation and achievable RV precision, (ii) a sensitivity analysis for 15 Zernike polynomials and how RV precision changes as a function of PSF phase error, and (iii) barycentric velocity shifts under the influence of spatially varying aberrations are presented in \S\ref{sec:results}. A summary and concluding remarks are provided in \S\ref{sec:summary}. To aid in the development of future instruments, the PSF aberration generation code from this study and spectrograph simulation code from \citenum{BechterE_19a}, has been made publicly available on Github.\footnote{\url{https://github.com/ebechter/InstrumentSimulator}}

\section{Methodology}\label{sec:method}

\begin{figure*}[t]
    \centering
    \includegraphics[width=\textwidth]{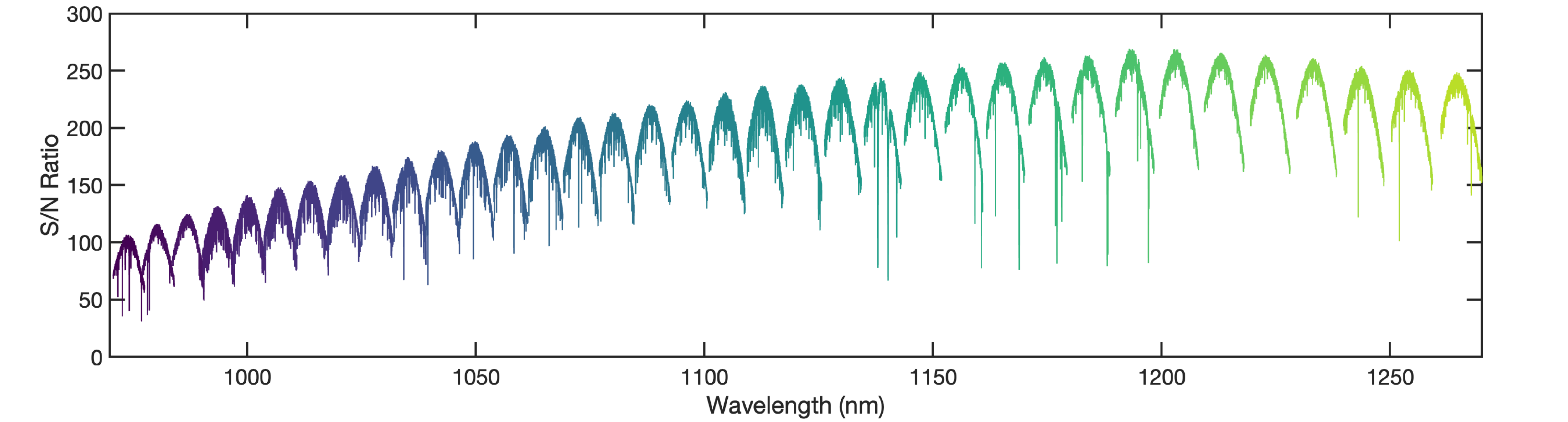}
    \caption[Simulated stellar spectrum used throughout analysis.]{Simulation of an M0V, $I=10$ star used to quantify RV precision as a function of aberration amplitude. A total of 36 spectral traces span grating orders $m = 117-152$.}
    \label{fig:input_spectrum}
\end{figure*}

Numerical simulations used to characterize the impact of optical aberrations draw from a foundation of code developed to model the iLocater spectrograph \cite{BechterE_19a}. Used previously to construct an instrument RV error budget\cite{BechterA_18}, and most recently to quantify uncertainties expected from NIR hybrid array HxRG detectors,\cite{BechterE_19b} this code basis has been augmented to incorporate the effects of wavefront phase errors.

The simulator code includes a variety of user-inputs to synthesise a complete observing scene including: zenith angle, seeing conditions, synthetic stellar catalog, the LBT adaptive optics system, wavelength-dependent telescope and spectrograph throughput, and polarization effects. For the present analysis, most extraneous settings were disabled, allowing for instrument PSF aberrations to be studied in isolation. Since the analysis focuses on stellar RV measurements, the same synthetic stellar spectrum has been used consistently throughout, against which the impact of instrument PSF variations is measured. 

An M0V star with apparent magnitude $I = 10$ is used as a representative observing target. Figure~\ref{fig:input_spectrum} shows the signal-to-noise ratio (SNR) of the simulated star for photon-noise-limited observations at iLocater's spectral resolution ($\rm R=132,000-273,000$). Atmospheric contamination from the sky-background and telluric absorption lines are not included in the simulations which would typically dominate central spectral orders ($\rm \lambda = 1.1-1.2\mu m$) located between the astronomical Y-band and J-band.

\begin{figure*}[t]
\begin{center}
\includegraphics[width=0.9\textwidth]{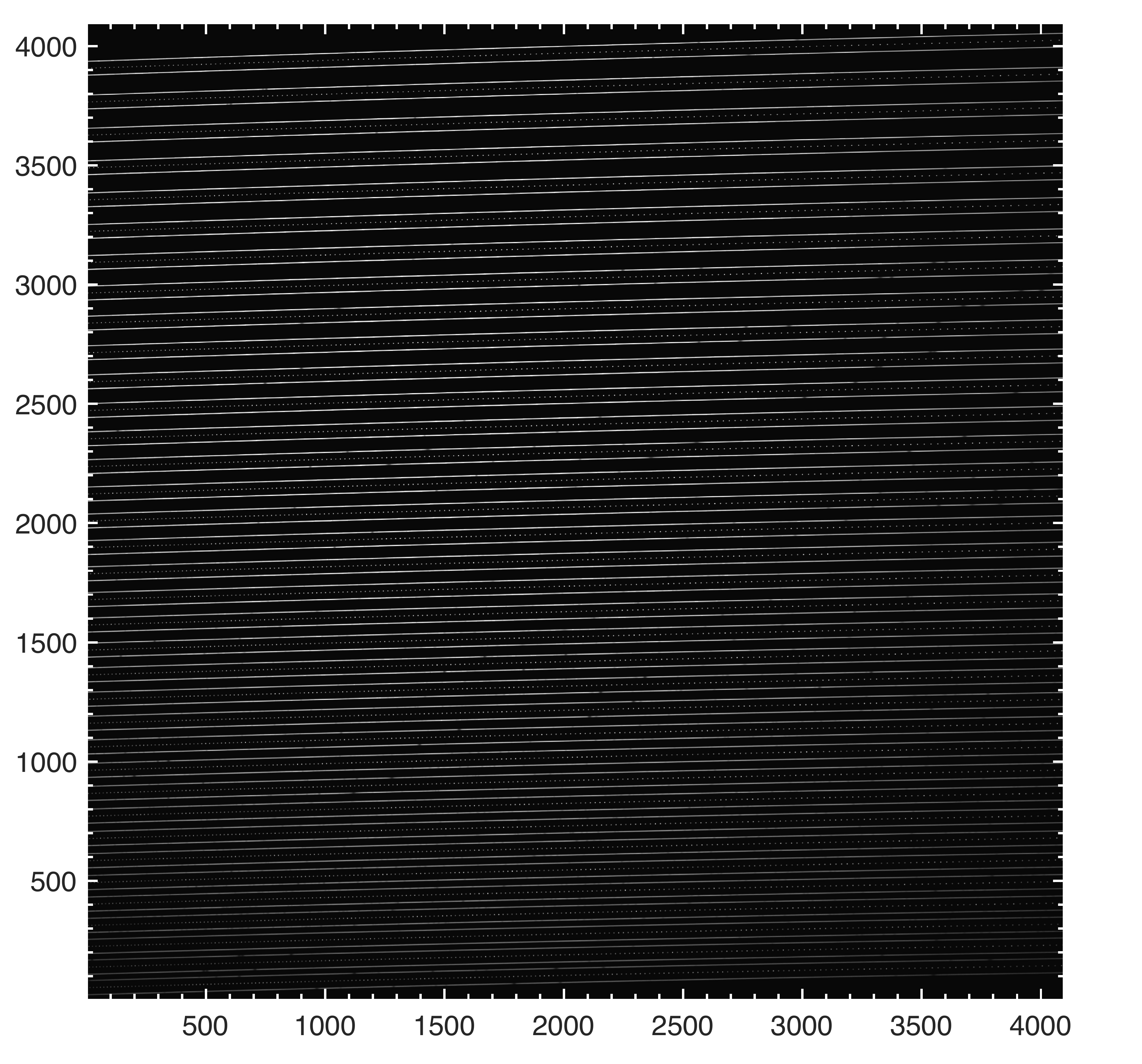}
\end{center}
\caption[Full frame simulation image]{ \label{fig:fullframe} Simulated noiseless focal plane using iLocater's optical design. Axes are bounded using the physical footprint of a 10$\mu m$ pixel H4RG-10 detector. Such frames are generated to study varying aberration types and magnitudes on RV precision.}
\end{figure*} 


The simulator initially resamples the target stellar spectrum to the appropriate resolution. In the case of an unaberrated beam, flux samples along a spectral order are convolved with a single PSF kernel to construct each order. For an aberrated beam, a variable wave-front error vector (\S\ref{sec:PSF}) is introduced prior to convolution to modify the PSF accordingly. This process allows for spatially varying PSF's across the dispersion direction of the array and for each spectral order. Spatial and spectral solutions ($x$, $y$, $\lambda$) for iLocater's optical design are derived from Zemax models and used to map the incidence of starlight and calibration light onto the focal plane array.   

Figure~\ref{fig:fullframe} shows an example simulation product that is free of aberrations. Top and bottom traces in each spectral order contain science light from each primary mirror of the LBT, while the central trace displays a 10 GHz etalon spectrum used for wavelength calibration. The 36 spectral orders span $m = 117-152$ from top to bottom. The dispersion direction runs horizontally on the page with wavelengths increasing from left to right. Cross-dispersion is in the vertical direction.

iLocater's data reduction pipeline is used to locate and extract spectral orders, apply a wavelength solution, and cross-correlate extracted spectra \cite{BechterE_19a}. A custom built binary mask based on M0V stars is used to derive wavelength shifts and compute a RV shift. A simulation free of aberrations is recorded as a benchmark. Aberrations are introduced and compared to the benchmark using otherwise identical settings. 

\subsection{PSF Generation}\label{sec:PSF}
iLocater's acquisition camera steers and couples each independent beam from the LBT's binocular primary mirrors into SMFs\cite{Crass_20}. The spectrograph is then illuminated by these SMFs which have a spatial output profile that can be closely approximated by a circular Gaussian function. The spectrograph optics are designed to preserve this mode profile by minimizing aberrations and diffraction from edge effects. As such, each optical element is over-sized to approximately twice the diameter of the Gaussian beam ($\mathrm{1/e^2}$ intensity) to maintain the Gaussian profile through the full optical path \cite{Robertson_12}.

\begin{figure}[t]
 \begin{center}
 \begin{tabular}{c} 
 \includegraphics[width=\textwidth]{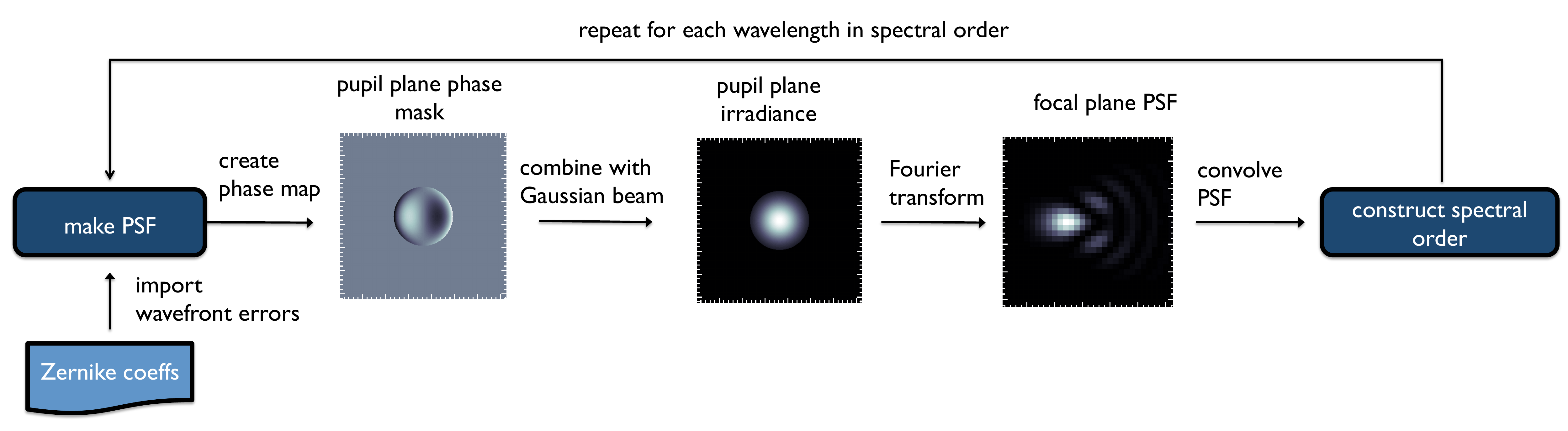}
\end{tabular}
 \end{center}
 \caption[PSF generation schematic] 
 {\label{fig:makePSF} PSF generation schematic. Zernike coefficients are provided for each sample along the spectral order, corresponding to the phase error at that wavelength. The PSF at the focal plane is generated by a single step Fourier transform of the complex pupil.}
 \end{figure}  

Figure~\ref{fig:makePSF} provides an overview of the steps involved in the aberrated PSF generation process. An unaberrated Gaussian PSF from the focal plane is numerically propagated to an arbitrary pupil plane where phase aberrations are applied. Aberrations are generated by modifying the electric field phase based on the shape of the desired distortion. The wavefront, $W(x,y)$, is defined as the sum over Zernike aberration terms:   
\begin{equation}
   W(x,y) = \sum_{n=0}^{\infty}\sum_{m=0}^{n}z_{nm}N^{m}_{n}P(x,y)_{n}^{m}, 
\end{equation}
where, $z$, $N$, and $P(x,y)$ are the Zernike expansion coefficient, normalization factor, and orthogonal circle polynomials respectively. User-supplied coefficients represent Zernike term amplitudes up to the first 16 Zernike polynomials following the Wyant\footnote{The choice for Wyant/Fringe indexing was made to follow typical optical software and lens design. There are many higher order polynomials beyond the first 16 which have not been explored in this paper and which should also be considered as part of an instrument design.} indexing convention\cite{Wyant_06}.

Zernike polynomials are applied to the Gaussian beam in a pupil plane, where the polynomial surface is referenced to the $\mathrm{1/e^2}$ diameter. This is chosen to ensure the spatial frequencies in each Zernike term and the RMS wavefront errors overlap with the majority of the optical beam. The Gaussian beam is truncated numerically and combined with the phase map, resulting in a complex pupil function, $E(x,y)$:
\begin{equation}
\label{pupil equation}
    E(x,y) = A(x,y)e^{i W(\lambda,x,y)}
\end{equation}
where $A(x,y)$ is the truncated Gaussian amplitude, representing the physical Gaussian beam truncation due to finite size of optical elements. Zernike coefficients and amplitudes are individually tuned for each wavelength in $W(\lambda,x,y)$. The PSF is then generated by taking the square modulus of the Fourier transform of the pupil function.

\begin{figure*}[t]
\begin{center}
\includegraphics[width=\textwidth]{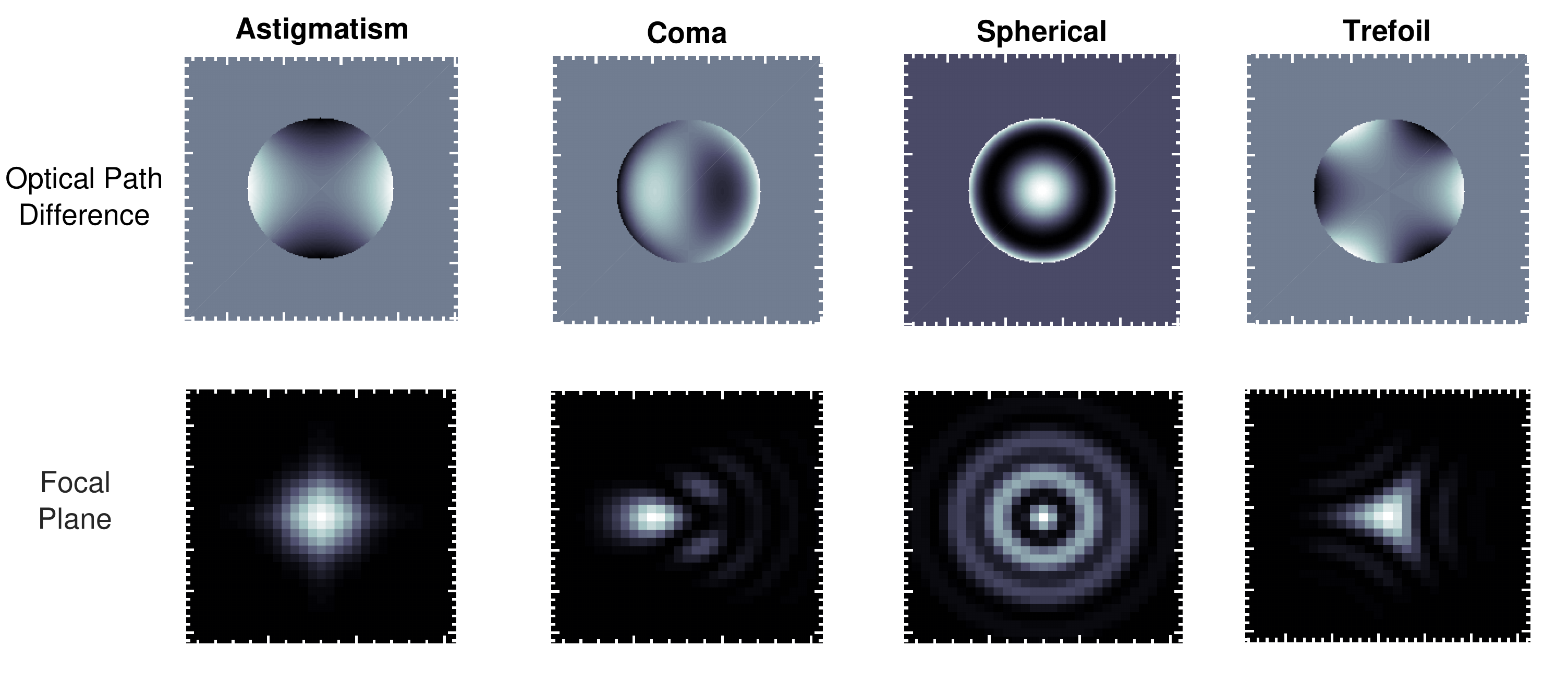}
\end{center}
\caption[Example Aberrations]{ \label{fig:example_zernike} Representative simulated optical aberrations. (\textbf{Top Row}:) Phase maps represented by the Zernike circle polynomials (primary astigmatism, primary coma, primary spherical, and primary trefoil). (\textbf{Bottom Row}:) Focal plane instrument PSFs after combining the phase map with a Gaussian beam. Instrument PSF kernels are used via convolution to map spectral point-intensities to detector coordinates.}
\end{figure*} 

Figure~\ref{fig:example_zernike} shows an example of four common optical aberrations (astigmatism, coma, spherical, and trefoil) with their respective phase perturbations and focal plane result. We qualitatively describe the resulting change in PSF shape as either: (i) errors with horizontally symmetric geometries that enlarge the PSF and degrade spectral resolution; or (ii) errors with asymmetric geometries along the dispersion axis that cause a shift in spectral line center. The latter type of distribution imitates a Doppler shift and degrades RV precision. Using this terminology, the aberrations in Figure~\ref{fig:example_zernike} would be classified from left to right as: (astigmatism) symmetric, (coma) asymmetric, (spherical) symmetric, and (trefoil) asymmetric, respectively. 

\subsection{Description of Simulations}
We use the simulation components outlined previously to assess the impact of optical aberrations on RV precision in three different ways. First, the Doppler content of spectra are studied both prior to and after introducing aberrations (\S\ref{sec:dopp-loss method}). This methodology captures the effects of resolution degradation on photon-noise-limited RV precision. Second, cross-correlation algorithms are used to measure centroid shifts in aberrated stellar lines (\S\ref{sec:line-shift method}) and calibration light (\S\ref{sec:calibration_method}). Of particular concern are aberrations that displace the photo-center of the instrument PSF in the dispersion direction as these mimic motion induced by an astronomical Doppler shift. Finally, the effect of Earth's barycentric motion is considered in the presence of spatially varying aberrations across the focal plane (\S\ref{sec:spatial-var method}). Sections \ref{sec:dopp-loss method}-\ref{sec:spatial-var method} describe the methodology behind each simulation. 

\subsubsection{Doppler Content}
\label{sec:dopp-loss method}

We use photon-noise-limited RV precision, $\sigma_{\rm ph}$, as a metric to quantify Doppler content,
\begin{equation}
\sigma_{\rm ph}=\frac{1}{\sqrt{\sum \left(\frac{dI/dV}{\epsilon_I} \right)^2}},
\label{eq:dopplerquality}
\end{equation}
where $dI/dV$ represents the slope of the measured stellar intensity as a function of wavelength (expressed in velocity units) and $\epsilon_I=\sqrt{N_{\rm ph}}/N_{\rm ph}$ is the fractional Poisson error\cite{Butler_96}. The majority of Doppler content is contained in the ``wings'' of stellar absorption lines; a line with a shallow gradient contains less Doppler information than a line with a steep gradient\cite{Fischer_16}. 

Aberrations displace energy from the instrument PSF core and act as a broadening-kernel that diminishes the slope of absorption lines. Aberrations that enlarge the instrument PSF, but do not drastically alter PSF morphology (e.g. spherical), still reduce the precision of a Doppler spectrometer. Since $dI/dV$ only acts as a measure of relative line slope, aberrations that broaden the PSF while also shifting the center of light distribution in the dispersion direction are not fully captured by Equation~\ref{eq:dopplerquality}. PSF shapes that shift photo-centers are discussed in \S\ref{sec:line-shift method}.

We simulate Doppler information loss by injecting aberrated PSFs into the simulation process following the procedure in \S\ref{sec:PSF}, using the M0V star in Figure~\ref{fig:input_spectrum} as the test spectrum. Aberration amplitudes are varied up to 0.5 waves of phase (RMS). Equation~\ref{eq:dopplerquality} is used to quantify the change in Doppler content in each simulation.

\subsubsection{Absorption Line Shifts}
\label{sec:line-shift method}

A common algorithm used for recovering Doppler velocities from stellar spectra is masked cross-correlation\cite{baranne_96}. We implement our own cross-correlation method, described in a previous publication\cite{BechterE_19a} to quantify the effect of aberrations on cross-correlation algorithms. Similar to the procedure in \S\ref{sec:dopp-loss method}, the effect of 15 common aberrations (the first 16 Zernikes not including piston) are simulated individually and applied to the stellar spectrum. RVs are measured from each modified stellar spectrum by applying the same binary mask in each case. After subtracting off the mask-offset velocity, the remaining residual velocity captures the effect of aberrations modifying individual absorption line profiles. 

Although real-world instrument PSF's are comprised of many Zernike modes, we simulate the effect of each aberration on the cross-correlation method individually to determine which aberrations are most detrimental. For each aberration, phase errors are modified following the process in \S\ref{sec:PSF}, using the MOV star in Figure~\ref{fig:input_spectrum} as the test spectrum. Aberration phase errors range from 0.05 to 0.5 waves, in increments of 0.05 waves. Results are explored in \S\ref{sec:sensstudy}.

\subsubsection{Tracking Line Drifts with Calibration Sources}
\label{sec:calibration_method}

Further, we include the effect of aberrations in the wavelength calibration system. Prior to cross-correlation between binary mask and the stellar spectrum, a wavelength solution is derived using calibration images\cite{BechterE_19a}. Aberrations that act to shift the photo-center of light along the dispersion direction should induce a displacement of comparable magnitude in the spectrum used for wavelength calibration. In other words, the wavelength solution would be shifted in velocity space in a similar magnitude to the results of cross-correlation, greatly mitigating the effect studied in \S\ref{sec:line-shift method} and \S\ref{sec:sensstudy}. 

The calibration source is studied using identical conditions to those described in \S\ref{sec:line-shift method}. A binary mask is tailored for the 10GHz calibration spectrum for use in cross-correlation. This enables a direct comparison of how stellar and calibration spectra respond under the effects of optical aberrations. Subtracting the calibration source RV from the stellar RV at each simulated aberration and magnitude gives a residual velocity. This residual velocity captures the aberration-induced velocity that calibration sources are unable to track. 

\begin{figure}[t]
\begin{center}
\begin{tabular}{c} 
\includegraphics[width=\textwidth]{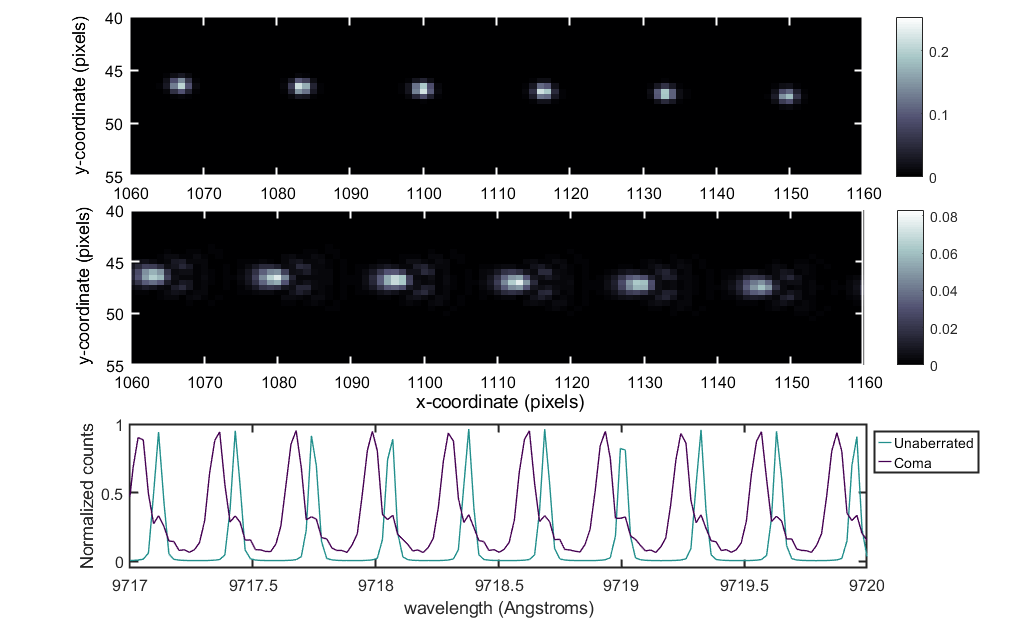}
\end{tabular}
\end{center}
\caption[Example of coma on the detector face] 
{Effect of coma on spectral peaks of 10 GHz etalon used as a wavelength calibration source.}
\label{fig:etalon_comparison}
\end{figure}

\begin{figure}
\begin{center}
\begin{tabular}{c} 
\includegraphics[width=0.6\textwidth]{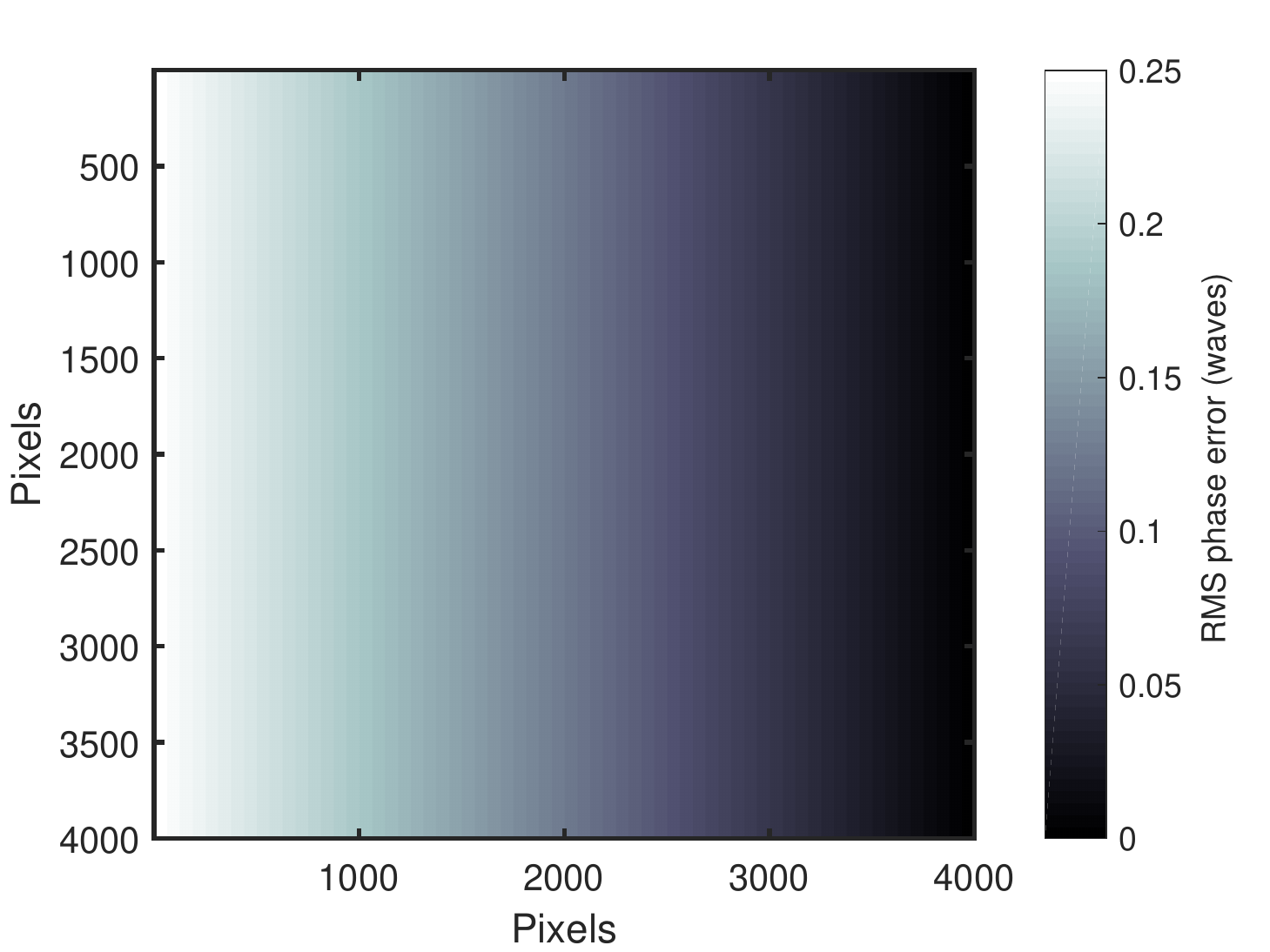}
\end{tabular}
\end{center}
\caption[Aberration amplitude map] 
{ \label{fig:abbmap} Aberration amplitude map matching the dimensions of the 4k x 4k detector. In this example, a linear gradient is applied from 0.25 waves to 0 going left to right, affecting all orders equally.}
\end{figure}

\subsubsection{Spatial Variations Across Array}
\label{sec:spatial-var method}

The conditions under which stellar spectra and calibration spectra are recorded are not strictly identical. Even in the case that scientific and calibration spectra are measured contemporaneously using two separate fibers, differential aberrations may still be present between different instrument input fibers as they take slightly different optical paths. Alternatively, if scientific and calibration spectra are measured using the same fiber, aberrations may evolve in time between these images due to instrument thermal or environmental changes.

Even with a perfectly stable instrument, Earth's barycentric motion causes stellar spectra to shift by dozens of pixels within a specific order over the course of an observing season, while calibration spectra do not experience such translation. This effect introduces differential aberrations between stellar and calibration spectra. As the cross-correlation algorithm relies on stability for a precise RV measurement, differential aberrations introduce an additional source of RV uncertainty.
 
To quantify the effects of spatially varying aberrations, we generate a spatial aberration map that follows a linear, horizontal gradient with respect to the detector plane (Figure~\ref{fig:abbmap}). While each spectrometer has a unique and complex spatial aberration pattern, Figure~\ref{fig:abbmap} provides a representative test case to demonstrate the principal effect. Each point within the heat-map represents the RMS phase error applied to the instrument PSF prior to convolution. Spectral data on the left side of the detector experience a higher magnitude of a particular aberration compared to points on the right side. Each aberration type is simulated making use of the aberration map in Figure~\ref{fig:abbmap}. Barycentric velocities are then introduced, ranging from -30 km/s to +30 km/s. Cross correlation methods are used to quantify RV residuals at each velocity offset. 

\section{Results}\label{sec:results}

\subsection{Doppler Content}
\label{sec:Dopplerinfo}

\begin{figure*}[t]
    \centering
    \includegraphics[trim=1.0cm 0cm 1.0cm 0.75cm,clip,width=\textwidth]{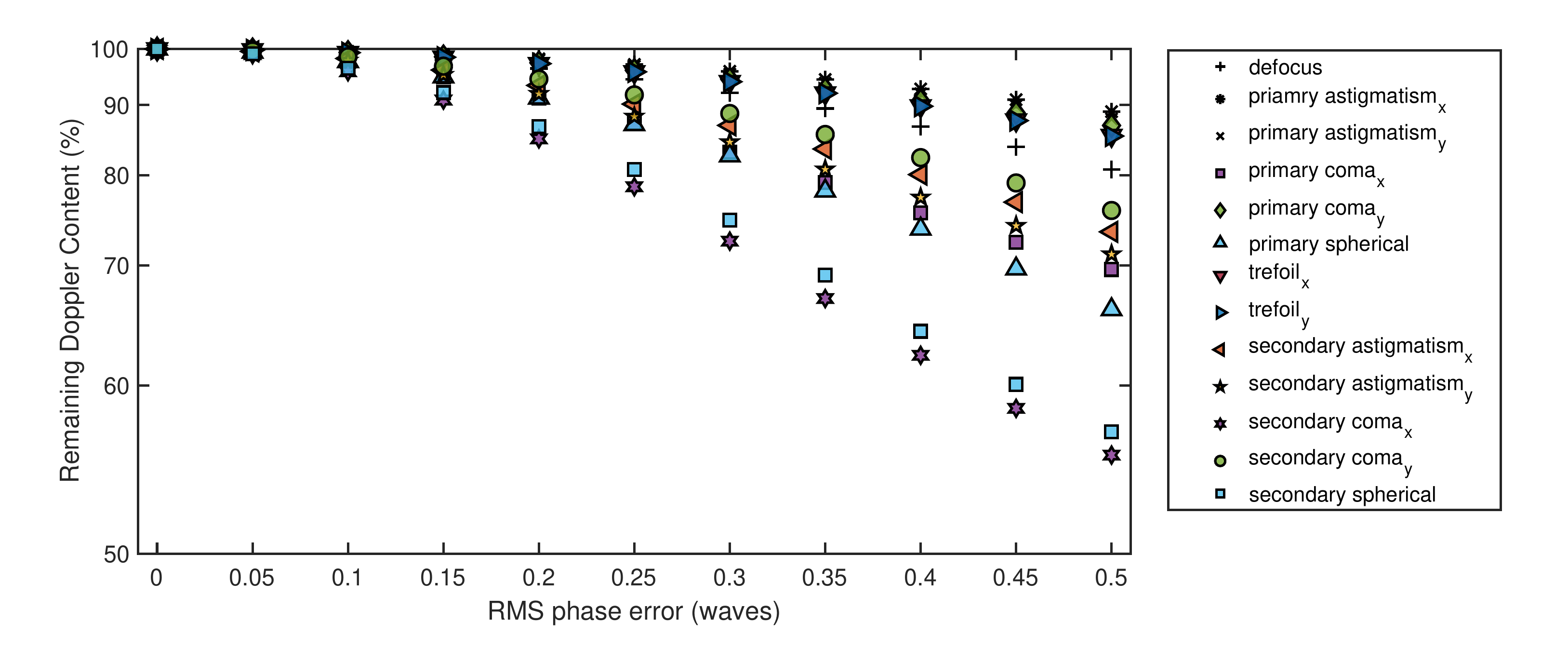}
    \caption[Doppler content as a function of phase error.] {Photon-noise-limited RV precision (Doppler content) relative to an unaberrated spectrum. Achievable RV precision is reduced as RMS phase error is increased. Aberrations that enlarge the PSF size substantially, such as primary and secondary coma and spherical, remove the most Doppler information from the stellar spectrum.}
    \label{fig:doppler_loss}
\end{figure*}

Figure~\ref{fig:doppler_loss} shows the rate at which Doppler content decreases as the severity of aberration increases in the system. Doppler content (y-axis), a measure of RV contained in a stellar spectrum, is calculated using Equation~\ref{eq:dopplerquality} and computed at a number of RMS phase errors shown on the x-axis. Because aberrations act to enlarge the instrument PSF, thereby reducing overall image quality, Doppler content decreases as phase errors increase. The results in  Figure~\ref{fig:doppler_loss} reveal that secondary spherical and secondary horizontal coma aberrations degrade Doppler content at a significantly faster rate than other aberrations. 

Setting thresholds on acceptable Doppler precision lost due to aberrations can yield upper limits on tolerable aberration magnitudes. For example, the effects of secondary coma at 0.2 waves of RMS phase error reduce Doppler precision below 90\% of the achievable precision of the same instrument without aberrations. Such limits can be informative for instrument builders and optical engineers during design and testing phases as many trade-offs will have to be considered, e.g. between instrument throughput, spectral resolution, and integration time.

\subsection{Absorption Line Shifts}
\label{sec:sensstudy}

\begin{figure}[t]
\begin{center}
\begin{tabular}{c} 
\includegraphics[trim=0.15cm 0cm 0.5cm 0.75cm,clip,width=\textwidth]{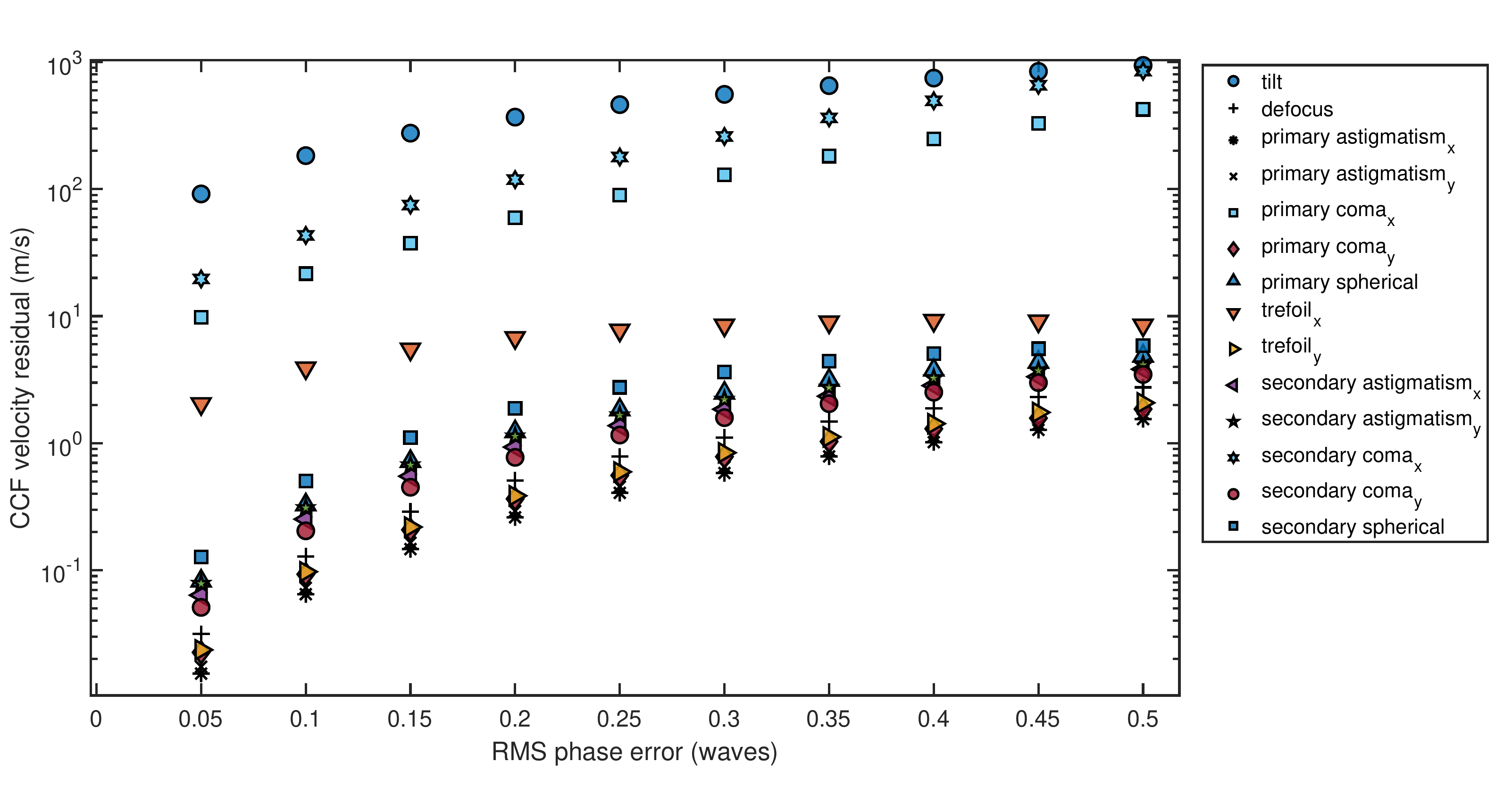}
\end{tabular}
\end{center}
\caption[Absorption line shifts.]
{ \label{fig:sensstudy}  CCF velocity residuals as a function of RMS phase error. As phase error increases, the degree to which absorption line profiles are modified and displaced also increases, resulting in greater CCF velocity residuals.}
\end{figure}

Figure~\ref{fig:sensstudy} shows how rapidly aberrations increase the displacement of the stellar spectrum along the dispersion direction\footnote{$\rm Piston$ and $\rm tilt_y$ (tilt acting perpendicular to the dispersion direction) have zero and nearly zero impact on RV and are omitted from the figure and following analysis.}. The velocity residuals, as measured by cross-correlation, are shown on the y-axis using the unaberrated stellar spectrum as the relative measure. As phase error increases for each aberration, absorption lines become further displaced causing cross-correlation residuals to increase. 

The cross-correlation technique is sensitive to aberrations that induce PSF photo-center displacements along the spectrograph dispersion axis. Due to order curvature on the detector, there is a minor coupling between the two coordinate systems, such that aberrations marked with a ``y'' still have a slight non-zero impact as measured by cross-correlation. While $\mathrm{tilt}$ appears to be one of the most damaging aberrations, it is well-corrected by simultaneous wavelength calibration (explored in \S\ref{sec:calibration}). We find that primary and secondary $\rm coma_x$ displace the stellar spectrum on the detector on the order of tens to hundreds of ms$^{-1}$ for each $1/20$th of an RMS phase error increment. $\rm Trefoil_x$ also shows greater than 1~ms$^{-1}$ residual at only 0.05 RMS phase error. 
Phase errors that do not result in dispersion-direction shifted stellar lines, e.g. astigmatism, defocus, and spherical, are less impactful to cross-correlation methods. While the PSF shape becomes enlarged and distorted as light is displaced from the core, the Doppler photo-center remains constant. In such cases, the non-zero velocity residual in Figure~\ref{fig:sensstudy} is caused by a combination of order curvature (noted above) and a reduction in spectral resolution, which reduces the precision of the cross-correlation fitting process, analogous to results in \S\ref{sec:Dopplerinfo}. Horizontal coma (primary and secondary) and trefoil can pose significant risk to precision RV instruments at only $1/20$th of a wave.

\subsection{Tracking Line Drifts with Calibration Sources}
\label{sec:calibration}

Examining Figures~\ref{fig:doppler_loss} and~\ref{fig:sensstudy}, it is clear that primary and secondary aberrations tend to produce qualitatively similar results. Also, of the aberrations having distinct $x$- and $y$-components, the $x$-component (dispersion direction) contributes most to measured Doppler error. To condense the remaining analysis, we only consider a subset of the initial 16 Zernike terms.

\begin{figure}[t]
\begin{center}
\begin{tabular}{c} 
\includegraphics[trim=0.75cm 0.5cm 1.0cm 0.25cm,clip,width=\textwidth]{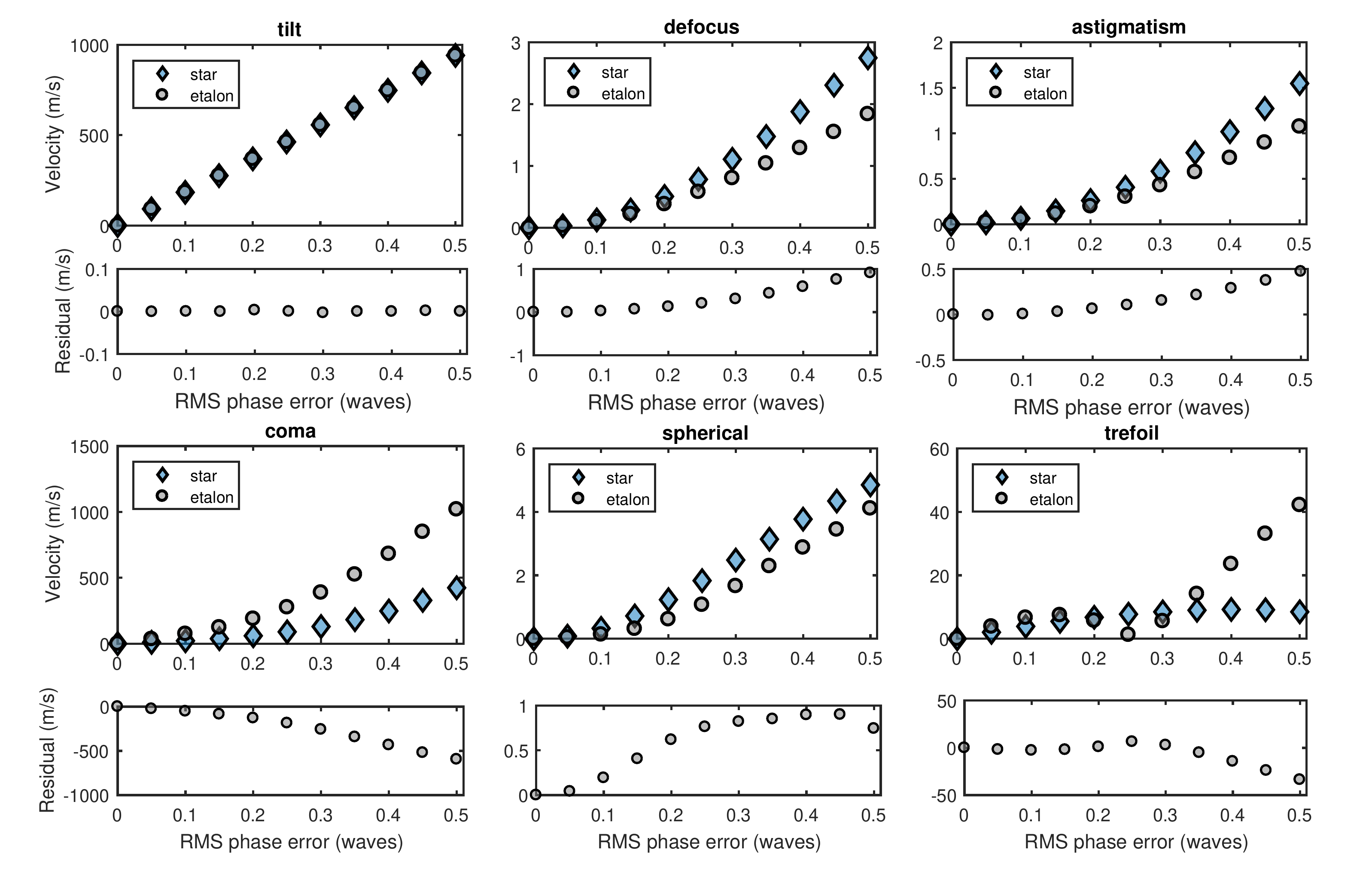}
\end{tabular}
\end{center}
\caption[Correcting line shifts with calibration spectra.] 
{ \label{fig:calib_residuals} In each panel, a stellar spectrum and wavelength calibration spectrum are simulated using the same magnitude and type of aberration. Residuals shown below show the difference between star and etalon velocities, quantifying how each aberration affects stellar and calibration spectra in different ways.}
\end{figure} 

Figure~\ref{fig:calib_residuals} quantifies how precisely a wavelength calibration unit is able to track and remove the effect described in the previous section {\S\ref{sec:sensstudy}}. Velocity residuals are measured on the y-axis at RMS phase errors ranging from 0.1 to 0.5 waves for 6 primary aberrations. Below each panel shows the difference between the star and etalon velocity residuals, representing the net impact of each aberration on RV measurements. 

Tilt is nearly perfectly compensated, only varying due to algorithmic or simulation precision errors. Defocus and astigmatism are well-corrected at 0.1 waves of RMS phase error and below ($\rm \leq 0.05~ms^{-1}$). However, in the case of coma and trefoil, a very large offset occurs. This result is in agreement with those presented in \S\ref{sec:Dopplerinfo} and \S\ref{sec:sensstudy}, where coma is also found to be the most harmful aberration to achieving precise RV measurements. Instrument builders should direct their attention to the residuals panel for each aberration. This panel provides the necessary details to set thresholds on individual aberrations as well as a guide to use when considering how certain aberrations impact the calibration spectrum differently than the stellar spectrum (e.g. trefoil).

\subsection{Spatial Variations}
\label{sec:spatial}

\begin{figure}[t!]
\begin{center}
\begin{tabular}{c} 
\includegraphics[width=1\textwidth]{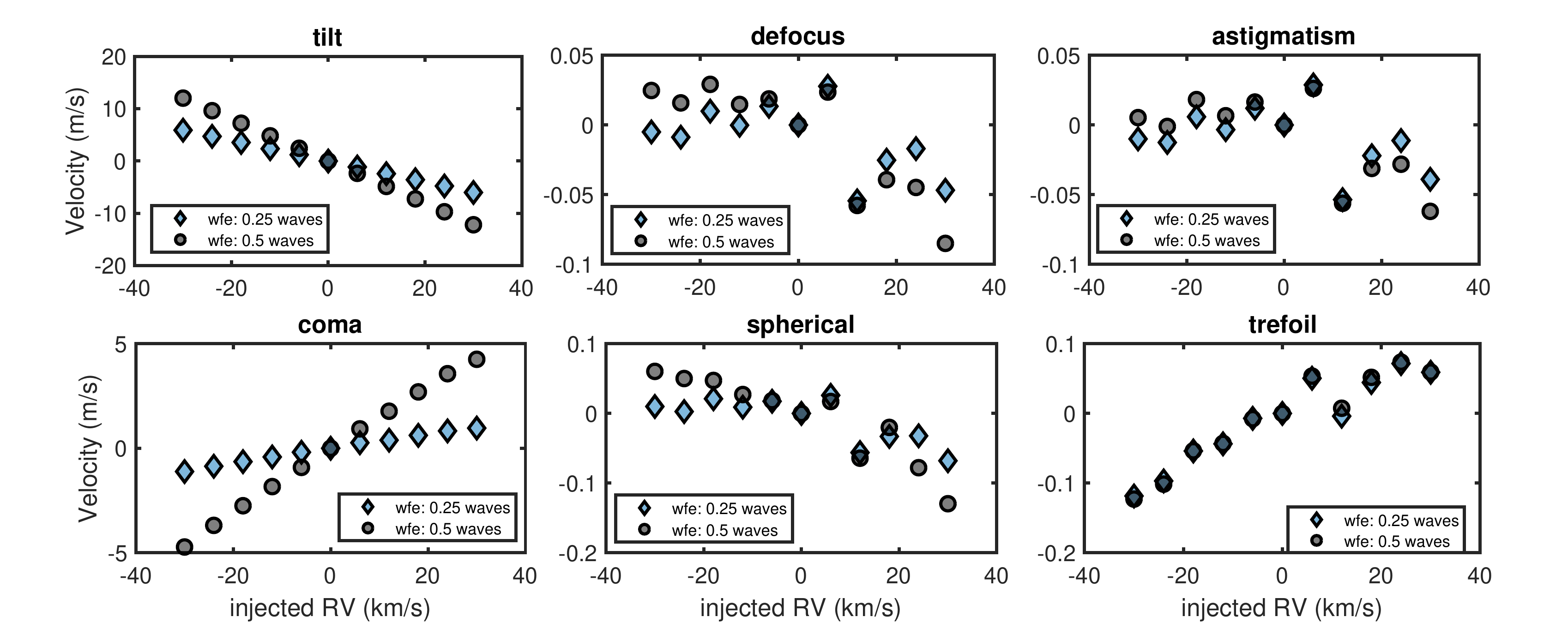}
\end{tabular}
\end{center}
\caption[Map results] 
{ \label{fig:mapresults} Residual RV evaluated for barycentric motions between -30 and 30 kms$^{-1}$. Aberration types are considered for two different linear gradients, ranging from 0.25-0.00 waves and 0.50-0.00 waves, similar to the gradient shown in Figure~\ref{fig:abbmap}.}
\end{figure}

Figure~\ref{fig:mapresults} shows how spatially varying aberrations impact RV measurements, quantified using cross-correlation methods. Recovered residual velocity from cross-correlation is indicated on the y-axis and injected Doppler velocity is on the x-axis, ranging from -30 to 30 km/s, approximating the range of Earth's barycentric velocity. Simulation results of each spatially varying phase map are indicated by circular and diamond markers, using a spatial map that contains linear gradients similar to Figure~\ref{fig:abbmap}. Figure~\ref{fig:mapresults} reveals that as the stellar spectrum is shifted due to changing barycentric velocity, causing spectral lines to be spatially displaced on the detector, the error in cross correlation measurement increases. 

Wavelength calibration units are at rest relative to the Doppler spectrograph, therefore there is no induced barycentric velocity on calibration spectra. Because of this, there is a velocity offset that changes over time between calibration spectra and observed stellar spectra, requiring a barycentric correction factor to be calculated at each observation. Because barycentric velocities can be corrected to the level of 1 cms$^{-1}$ in data reduction pipelines\cite{Wright_14}, our results in Figure~\ref{fig:mapresults} assume perfect barycentric correction. While barycentric calculations are able to provide a correction factor in velocity space, the shifted stellar spectrum still strikes the detector at a different location. Therefore, the residual velocity on the y-axis of Figure~\ref{fig:mapresults} should be interpreted as the remaining velocity error due to spatially-varying aberrations present in the system. 

As the injected barycentric RV increases, representing measurements at different observing seasons, each absorption line experiences a variation of aberration magnitude, adding a differential shift in stellar absorption lines. This effect is magnified when more extreme spatial variations are simulated, as shown by the increased differential values between the diamond and circle markers in Figure~\ref{fig:mapresults}. The results for coma and tilt aberrations show velocity residuals an order of magnitude greater than the other aberrations. However, in the case of tilt, because there is no change in PSF morphology, the velocity offset can be calibrated and compensated for using software and the wavelength calibration unit. Differential focus, astigmatism, and spherical aberrations act primarily to increase PSF size, reducing RV precision by reducing resolution, but without shifting stellar absorption line centers. Trefoil, in particular, shows a very small variation in RV when the slope of the aberration magnitude is increased. Coma aberrations present the highest risk to precision RV measurements, consistent with results from previous sections (\S\ref{sec:Dopplerinfo} and \S\ref{sec:calibration}).

\subsection{Summary of Results}
\label{summary_of_results}

\begin{table}[!htbp]
    \centering
    \begin{threeparttable}
\begin{tabulary}{.5\linewidth}{lcc}
        \toprule
			  Aberration~~~~~~  & ~~~~~~Doppler Content$^{[1]}$ ~~~~~~ &~~~~~~CCF errors$^{[2]}$~~~~~~\\ 
        \midrule          
                Tilt & $^{[3]}$ & $^{[3]}$ \\[3pt]
                Defocus & $\leq0.22$ & $\leq 0.18$   \\[3pt]
                Astigmatism$_x$ & $\leq0.27$  & $\leq0.25$ \\[3pt]
                Coma$_x$ & $\leq0.12$ & $\ll 0.05^{[4]}$  \\[3pt]
                Spherical & $\leq0.11$ & $\leq 0.07$ \\[3pt]
                Trefoil$_x$ &$\leq0.25$ & $\ll 0.05^{[4]}$  \\
        \bottomrule  
    \end{tabulary}
    \begin{tablenotes}
        \footnotesize
        \item[1] Maximum RMS phase error to preserve $\approx 95\%$ of Doppler content (Figure~\ref{fig:doppler_loss}).
        \item[2] Maximum RMS phase error for $\rm \Delta RV = 0.1~ms^{-1} $ including calibration unit correction (Figure \ref{fig:calib_residuals}).
	 \item[3] Tilt has no impact on PSF morphology. 
	 \item[4] The smallest phase aberration simulated is 0.05 waves RMS. 
    \end{tablenotes}
\end{threeparttable}
    \vspace{4pt}
 	\caption[Summary of Results.]{\label{tab:SummaryOfAbResults1} Summary of results from Doppler content and CCF aberration analyses. Maximum allowable wavefront phase errors are listed for each aberration in units of waves RMS.}
\end{table}

\begin{table}[!htbp]
    \centering
    \begin{threeparttable}
\begin{tabulary}{.5\linewidth}{lcc}
        \toprule
        & \multicolumn{2}{c}{Maximum seasonal RV variation (ms$^{-1}$)} \\ 
			  Linear gradient:~~~~  & ~~~~ $6.1 \times 10^{-5}$ waves/pixel ~~~~ &  ~~~~$1.2 \times 10^{-4}$ waves/pixel~~~~~\\ 
        \midrule          
                Tilt & 12.10 & 24.22 \\[3pt]
                Defocus & 0.07 & 0.15  \\[3pt]
                Astigmatism$_x$ & 0.08 & 0.83\\[3pt]
                Coma$_x$ & 2.10 & 8.97  \\[3pt]
                Spherical & 0.09 & 0.19 \\[3pt]
                Trefoil$_x$ & 0.19 & 0.20  \\
        \bottomrule  
    \end{tabulary}
\end{threeparttable}
    \vspace{4pt}
 	\caption[Summary of Results.]{\label{tab:SummaryOfAbResults2} Summary of results from Barycentric analysis. Each column considers different (linear, horizontal) spatial gradients across the detector (see Figure~\ref{fig:abbmap}). Results are in units of ms$^{-1}$ and correspond to the maximum RV variation induced by $\pm30$ kms$^{-1}$ extrema Barycentric motion for each aberration type (see Figure~\ref{fig:mapresults}). See text for discussion.}
\end{table}

Table~\ref{tab:SummaryOfAbResults1} summarizes the results from the Doppler content analysis (\S\ref{sec:Dopplerinfo}) and  CCF analysis (\S\ref{sec:calibration}). For each analysis, we assign reasonable thresholds on aberration phase errors to demonstrate how an instrument can make use of the simulations results. The first column in Table~\ref{tab:SummaryOfAbResults1} shows the upper limit of phase errors that preserve at least $95\%$ of the original Doppler information. The second column marks the maximum phase error such that no more than $\rm 0.1~ms^{-1}$ are imparted by an individual aberration after wavelength calibration correction. For each aberration, the threshold on phase errors set by the CCF analysis is lower than that set by the Doppler content analysis. However, depending on the individual instrument RV error budget, this may not always be the case. Using the above criteria, an instrument builder can see clearly that coma, spherical, and trefoil aberrations have very tight requirements on allowable phase aberration amplitude. 

Spatial variation analysis results (\S\ref{sec:spatial}) are presented separately in Table~\ref{tab:SummaryOfAbResults2} as the interpretation is not as straight-forward as those in Table~\ref{tab:SummaryOfAbResults1}. Each column in the table marks the slope of the simulated linear gradient. For each aberration listed, the maximum RV variation is measured through the full range of barycentric velocities ($\pm30$ km/s). For example, allowing for horizontal coma to change linearly by 0.25 waves across the detector face (4096 pixels for iLocater), or at a rate of $2.5\times10^{-6}$ waves per pixel in the dispersion direction, creates a RV uncertainty of several meters-per-second. This error quantifies the RV uncertainty beyond the correcting ability of calibration units (\S\ref{sec:spatial}). Table~\ref{tab:SummaryOfAbResults2} indicates horizontal coma causes more than an order of magnitude greater RV error than other aberrations listed. Spectrograph designs must therefore optimize optical prescriptions to specifically suppress these aberrations. 


\section{Concluding Remarks}\label{sec:summary}

Numerical simulations are now used early-on in the instrument development process to construct error budgets, estimate on-sky performance, and inform design decisions. In the case of EPRV spectrographs, improving Doppler precision to below 1 meter-per-second necessitates more ambitious designs than previous generations.\footnote{\url{https://exoplanets.nasa.gov/internal_resources/1556/}} Modern instruments command broader wavelength coverage, higher spectral resolution, larger optics and detectors, off-axis designs, novel materials, multiple input fibers, or even multiple telescopes\cite{Davis_17, Jovanovic_16, Pinna_16}. The advanced needs of these instruments, combined with improved technologies, means that effects which negligibly impacted performance previously may severely limit capabilities. Optical aberrations are one such effect which become increasingly important for diffraction-limited EPRV instruments fed using SMFs. While SMF-fed spectrographs benefit from a slow focal ratio camera which minimizes induced aberrations, the fast focal ratio of collimating fore-optics poses challenges to maintaining diffraction-limited performance in design and fabrication. It is therefore important to understand which  aberrations dominate performance and if these are required to be mitigated in instrument designs or can be mitigated through calibration.



We have developed simulations that model the influence of wavefront distortions on RV precision. Most power in the spectrum of optical aberrations from lenses and mirrors is contained in the lowest spatial frequencies. Studying the first 16 orthogonal Zernike modes, we find that the shape of various aberrations impacts performance in qualitatively different ways. Rotationally symmetric aberrations, such as defocus and spherical, reduce spectrograph resolution and diminish available Doppler content (\S\ref{sec:Dopplerinfo}). Asymmetric aberrations that elongate images in the cross-dispersion direction have only a minor impact on recovered RV information; the effect is small, but non-zero, for spectral traces having geometric curvature across the focal plane (Figure~\ref{fig:sensstudy}). Of particular concern are asymmetric aberrations that cause a photo-center shift along the dispersion axis of the spectrograph.

Horizontal coma ($\rm Coma_x$) and horizontal trefoil ($\rm Trefoil_x$) can easily exceed the RV threshold of the entire error budget (\S\ref{summary_of_results}). Such aberrations may be encountered with fast optical systems, or introduced by misalignment of optical elements. Horizontal coma and horizontal trefoil must be driven to zero during the spectrograph design and construction process. 

Wavelength calibration systems are able to partially correct for changes in PSF morphology, but only insofar as calibration light follows the same optical path, capturing the same aberrations as starlight. Differential aberrations, experienced between off-axis field points or through thermal changes in time (in between calibration measurements), decompose into common Zernike modes and can contribute non-negligibly to Doppler error budgets. Even with a perfectly stable instrument, the inherent wavelength dependence of the relative strength of phase aberrations is unavoidable, creating a gradient in image quality along the spectrograph dispersion direction.  

Over the course of an observing season, barycentric motion of the Earth around the Sun forces comparisons between spectral features at different locations on the focal plane array. At the resolutions needed to sample stellar absorption lines ($R > 100,000$), changes of $\Delta RV = \pm 30$ km s$^{-1}$ translates to RV shifts of several dozen pixels. Light from a wavelength calibration unit is not subject to barycentric Doppler shifts and therefore may not follow the same optical path as starlight even when using the same fiber.

In summary, certain aberrations disproportionately affect the precision of Doppler measurements. Identifying the strongest contributors, such as ``horizontal primary coma,'' ``horizontal secondary coma,'' and ``horizontal trefoil,'' allows an optimization of optical designs and the minimization of asymmetric aberrations when optimizing the performance of RV spectrographs. The results can in turn be used to derive maximum separations between fibers, field of view limits, acceptable distortion levels, and variations across the array. Instrument builders should model these second-order effects to ensure a fully optimized instrument performance.

\section{Acknowledgements}
We thank David Aikens, Joaquin Mason, and David King for helpful conversations about iLocater's optical design that helped motivate the simulations presented in this study. JRC acknowledges support from NASA's Early Career program (NNX13AB03G) and the NSF CAREER Fellowship (NSF 1654125). We are grateful for support from the Potenziani family and the Wolfe family.

\bibliography{manuscript.bib}   
\bibliographystyle{spiejour.bst}   

\end{spacing}

\end{document}